\documentclass{rnote}
\usepackage{epsfig}

\begin{document}

\author{G. Kie{\ss}lich$^*$) ($a$), A. Wacker ($a$), 
E. Sch{\"o}ll ($a$), 
A. Nauen ($b$), F. Hohls ($b$), and R. J. Haug ($b$)}
\address{(a) Institut f{\"u}r Theoretische Physik, Technische Universit{\"a}t
Berlin, Hardenbergstr. 36, D-10623 Berlin, Germany\\
(b) Institut f{\"u}r Festk{\"o}rperphysik, Universit{\"a}t Hannover, 
Appelstr. 2, D-30167 Hannover, Germany\\
$^*$Corresponding author;  Fax: +49-(0)30-314-21130, kieslich@physik.tu-berlin.de}

\submitted{September 30, 2002}

\title{Shot Noise in Tunneling through a Quantum Dot Array}

\maketitle

\paragraph{Abstract}

The shot noise suppression in a sample containing a layer of self-assembled InAs 
quantum dots has been investigated experimentally and theoretically. The observation
of a non-monotonic dependence of the Fano factor on the bias voltage in a regime where
only few quantum dot ground states contribute to the tunneling current is analyzed 
by a master equation model. Under the assumption of tunneling through states without Coulomb interaction 
this behaviour can be qualitatively reproduced by an analytical expression.

\paragraph{Introduction}

Shot noise measurements provide a sensitive tool for probing transport 
properties of charged particles in mesoscopic systems, e.g. tunneling through 
semiconductor heterostructures, 
which are not available by conductance measurements alone. The dynamic correlations between
individual tunneling events can reveal details of the potential shape and  
the effects of electron-electron interaction \cite{BLA00}.
Up to frequencies $f$ of the inverse transit time of carriers the spectral power density of the 
current noise is frequency-independent: $S(f)\approx S(0)$. For an uncorrelated flow of electrons 
(Poisson statistics of individual tunneling events) this value 
is proportional to the elementary charge $e$ and the stationary current $I$: $S_P(0)=2eI$ 
\cite{SCH18}. A reduction
of this value refers to negative correlations in the current (sub-Poissonian noise), 
e.g. caused by the Pauli exclusion principle or repulsive Coulomb interactions, quantified by the
Fano factor $\alpha=S(0)/S_P\le 1$. In double-barrier resonant tunneling structures this value is 
given by the ratio of the tunneling rates $\Gamma_{E/C}$ of emitter/collector barrier 
respectively: $\alpha=(\Gamma_E^2+\Gamma_C^2)/(\Gamma_E+\Gamma_C)^2$ \cite{CHE91}. 
In \cite{HER93} a master equation
model was applied in order to describe double-barrier tunneling through a metallic quantum dot (QD) 
including the Coulomb interaction which results in the Coulomb Blockade effect. 
In the plateau regions of the Coulomb staircase the shot
noise was Poissonian whereas the Fano factor shows dips in the region of steps which is 
quantitativly confirmed in the experiment \cite{BIR95}. For a 
semiconductor QD the shot noise 
in the nonlinear bias regime was studied in \cite{WAN98a} by a  nonequilibrium Green's function 
technique. They find in the plateau region a suppressed Fano factor and at the 
steps an enhancement of $\alpha$ still below the Poissonian value $\alpha =1$. 

Here, we use a master equation model \cite{KIE02,KIE02a} in order to investigate the 
shot noise in tunneling through electrostatically coupled semiconductor QDs. The 
distinction between the effect of the Pauli exclusion principle and Coulomb interaction on 
the shot noise suppression will be shown. 
We apply the theoretical result to a combined  shot noise and current 
measurement of tunneling through a layer of self-assembled QDs \cite{NAU02}.

\paragraph{Experiment}

\begin{figure}[ht]
  \begin{center}
    \includegraphics[width=\textwidth]{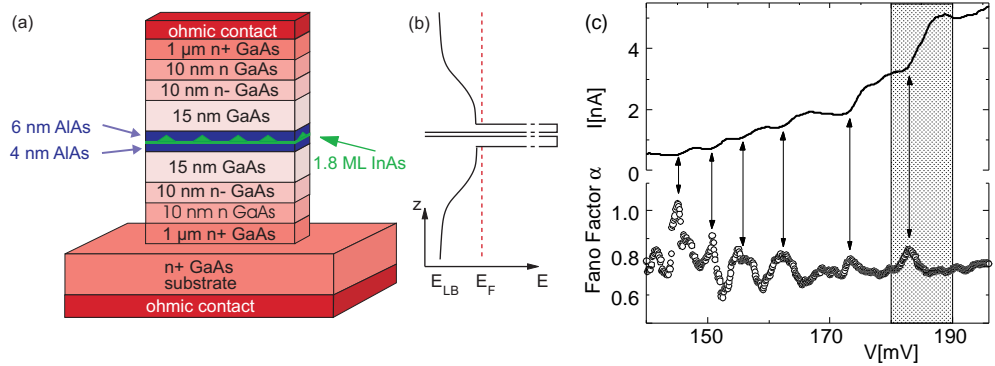}
    \caption{(a) Growth scheme of the sample. (b) The band structure of the sample and 
	Fermi energy 
	$E_F$. (c) upper trace: measured current-voltage characteristic; lower trace: Fano factor 
	$\alpha$ vs. bias voltage $V$, the noise data are smoothed using a seven point boxcar 
	averaging; Shaded region: compare Fig. \ref{fig2}b. Temperature: $T=$1.6 K.}
    \label{fig1}
  \end{center}
\end{figure}

A layer of self-assembled InAs QDs (10-15 nm diameter, 3 nm height \cite{HAP99}) is embedded in a 
GaAs-AlAs-GaAs
tunneling structure (Fig. \ref{fig1}a). The plane of QDs is sandwiched between two AlAs 
barriers of nominally 
4 nm and 6 nm thickness. About one million QDs are placed randomly on the area of an etched diode 
structure of 40$\mu$m$\times$40$\mu$m. A 15 nm undoped GaAs spacer layer and a GaAs buffer with 
graded doping on both sides of the resonant tunneling structure provide three-dimensional emitter and
collector contacts. Connection to the active layer is realized by annealed Au/Ge/Ni/Au contacts.
The respective conduction band edge and the Fermi energy $E_F=$13.6 meV for zero bias is depicted in 
Fig. \ref{fig1}b. 
By applying a bias voltage $V$ the current through the sample is measured (upper trace in Fig. 
\ref{fig1}c). The current-voltage characteristic shows steps which are assumed to be caused by
selection of resonant single QD ground states by single-electron tunneling \cite{HAP99,NAR97,ITS96}. 
The displayed bias direction corresponds to tunneling of electrons from the bottom to the top
of the pyramidal QDs. We have performed noise measurements in a frequency range from 0 to 100 kHz.
Above the cut-off frequency for $1/f$-noise of around 20 kHz we observe frequency independent noise
spectra. The temperature was 1.6 K.
The Fano factor $\alpha$ (lower trace in Fig. \ref{fig1}c) shows a average suppression 
$\alpha\approx$0.8 which can be used for the determination of the thickness of the collector 
barrier \cite{NAU02}. Furthermore, a non-monotonic behaviour of $\alpha$ is visible where the maxima 
correspond to steps in the current-voltage characteristic as indicated by arrows in 
Fig. \ref{fig1}c. This feature we address in the next section. 

\paragraph{Theory}

\begin{figure}[ht]
  \begin{center}
    \raisebox{0.5mm}{\includegraphics[scale=.5]{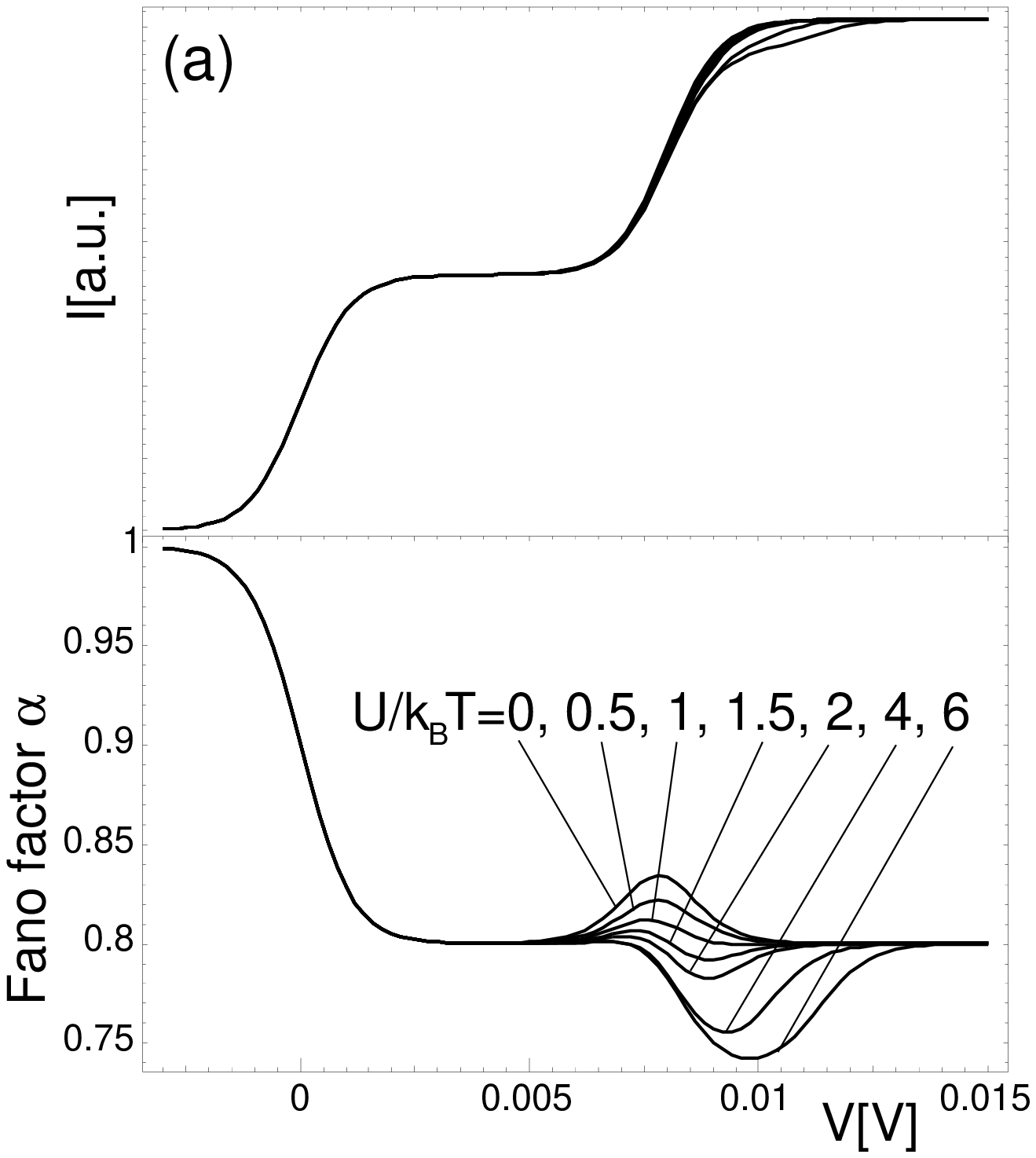}}
    \includegraphics[scale=.3]{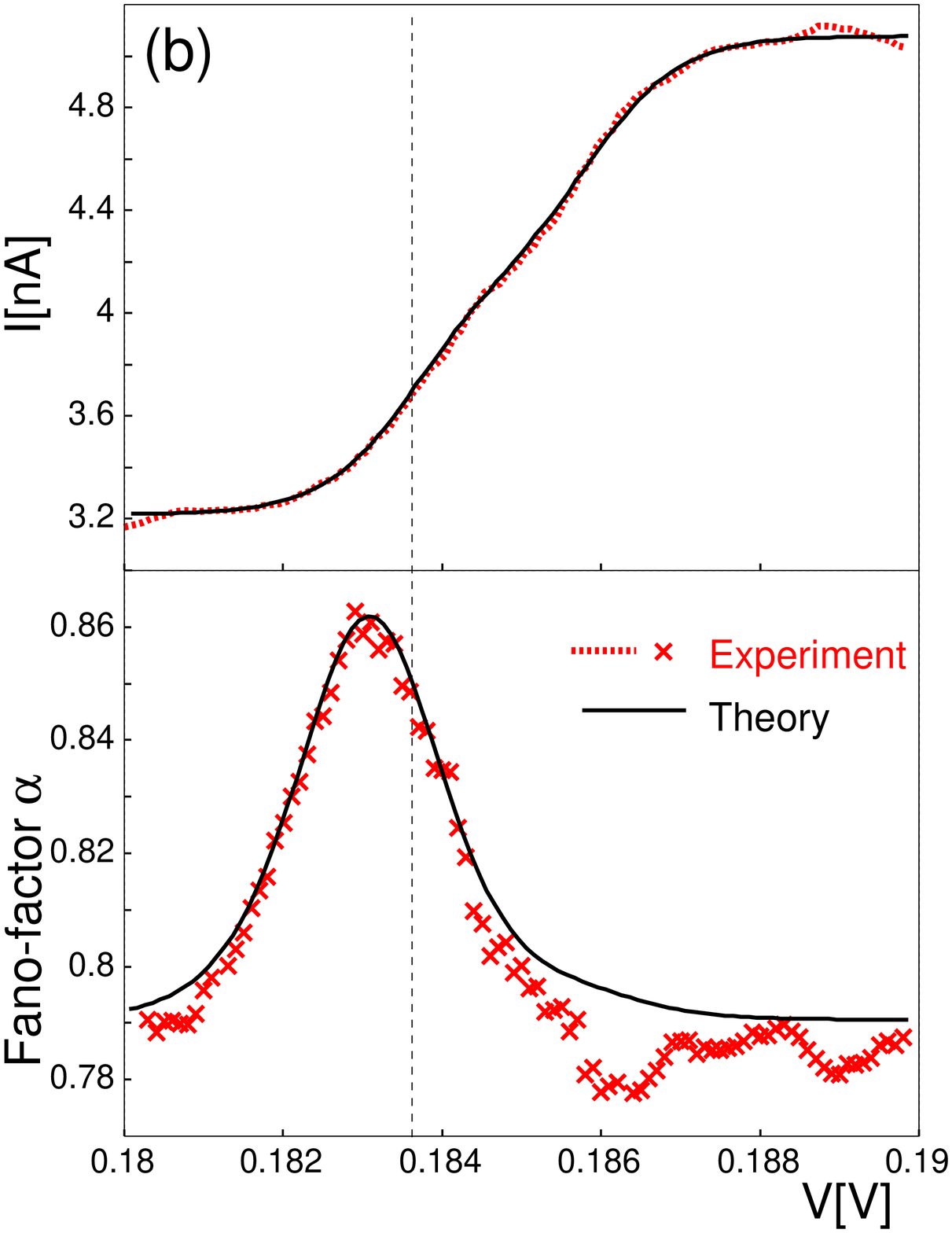}
    \caption{(a) upper trace: Calculated current-voltage characteristic (\ref{eq:current})
	for two QD states; lower trace: Corresponding Fano factor; $E_1=0$,
	$E_2=2$meV, $T=$1.6 K, $\eta =$0.3, $\gamma_i=$7.8 ($i=1,2$). (b) solid lines: 
	theory
	see text; dashed curve in upper picture and crosses in lower picture: experimental data, 
	see shaded region in 
	Fig. \ref{fig1}c}
    \label{fig2}
  \end{center}
\end{figure}

In order to calculate the Fano factor $\alpha$ and the stationary tunneling current $I$ 
through two Coulomb interacting non-degenerate QD states
we apply a master equation for the occupation probabilities: $\dot{\mathbf{P}}(t)=
\underline{\mathbf{M}}\cdot\mathbf{P}(t)$ with
$\mathbf{P}(t)=\left (P_{00}(t),P_{10}(t),P_{01}(t),P_{11}(t)\right )^T$  \cite{KIE02,KIE02a}.
In the matrix elements of $\underline{\mathbf{M}}$ the rates $\Gamma_{E/C}^i$ ($i=1,2$) for 
tunneling from the emitter/collector into the $i$-th QD state and vice versa enter.
The occupation of the contacts is treated in local equilibrium by Fermi functions
$f_E^i=(1+\exp{((E_i-\eta V)/(k_BT))})^{-1}$ and 
$f_E^{i,U}=(1+\exp{((E_i+U-\eta V)/(k_BT))})^{-1}$ ( with energy $E_i$ of $i$-th QD state, 
Coulomb interaction energy $U$, voltage drop $\eta$ across emitter barrier, 
and bias voltage $V$). We use $f_C^i=f_C^{i,U}=0$ assuming $eV\gg k_BT$.
In the derivation of an expression for the  spectral power density we follow 
the lines of Ref. \cite{HER93}:
For the time evolution of the occupation probabilities one defines the propagator
$\underline{\mathbf{T}}(t)\equiv\exp{(\underline{\mathbf{M}}t)}$
such that $\mathbf{P}(t)=\underline{\mathbf{T}}(t)\cdot\mathbf{P}(0)$.
The stationary occupation probability is obtained by $\underline{\mathbf{M}}\cdot\mathbf{P}^0=0$.
With the current operator for tunneling through the collector barrier 
$\underline{\mathbf{j}}_C$ the stationary current reads

\begin{equation}
I=\sum_i(\underline{\mathbf{j}}_C\cdot\mathbf{P}^0)_i.
\label{eq:current}
\end{equation}

The Fourier transform of the current auto-correlation function defines the frequency independent 
spectral power density: $S(0)=4\int_0^\infty dt\left (\langle i(t)i(0)\rangle -I^2\right )$.
This leads to the Fano factor

\begin{equation}
\alpha =\frac{S(0)}{2eI}
=1+\frac{2\int_0^\infty dt\left [\sum_i(\underline{\mathbf{j}}_C
\cdot\underline{\mathbf{T}}(t)
\cdot\underline{\mathbf{j}}_C\cdot\mathbf{P}^0)_i-I^2\right ]}{eI}
\label{eq:fano}
\end{equation}

In Fig. \ref{fig2}a the results of Eqs. (\ref{eq:current}) and (\ref{eq:fano}) for different 
Coulomb interaction strengths $U$ are shown. In the current (upper trace
of Fig. \ref{fig2}a) two steps occur due to the different resonances of the QD states with the 
emitter Fermi energy ($U=0$).  The corresponding Fano factor $\alpha$ (lower trace
of Fig. \ref{fig2}a) equals one below the first resonance. 
As a consequence of the Pauli exclusion principle the shot noise becomes suppressed
for tunneling through the non-interacting state $i$. The corresponding Fano factor is expressed by
$\alpha_i=1-\frac{2}{\gamma_i+2+1/\gamma_i}f_E^i$.
For biases below the second current step, only tunneling through state $i=1$ is possible and thus
$\alpha =\alpha_1$.
At the resonance of the second QD state the Fano factor has a peak which is also caused by the Pauli
exclusion principle: when the state $i=2$ becomes resonant with thermal excited electrons 
of the emitter contact an additional transport channel opens.
As $f_E^2\ll 1$, we find $\alpha_2\approx 1$, i.e. uncorrelated tunneling events close to the onset.
For tunneling through 
an arbitrary number of non-interacting QD states the current is $I=\sum_iI_i$ and we obtain 
the Fano factor

\begin{eqnarray}
\alpha=\frac{\sum_iI_i\alpha_i}{I}\quad\textrm{with}\quad I_i=\Gamma_E^i\left (\frac{1}{1+\gamma_i}
\right )f_E^i
\label{eq:Fano}
\end{eqnarray}

Thus $\alpha$ increases at the onset.
Further increasing the bias voltage enhances $f_E^2$ towards 1 and the Fano factor $\alpha_2$ decreases.
Consequently, $\alpha$ decreases as well.

With a finite Coulomb interaction $U$ a third step occurs in the current due to the resonance
of the double occupied state. For $U\approx k_BT$ there is hardly any difference in the current 
compared to the non-interacting case. But the Fano factor changes drastically: the peak
decreases and a dip arises caused by Coulomb correlations. Hence, even though the influence
of a small Coulomb interaction does not affect the current, the shot noise reacts very sensitively.

Now we apply the theoretical results to the experiment and consider the Fano factor peak in the
shaded region of Fig. \ref{fig1}c. Under the assumption of non-interacting QD states we fitted
the current step (upper trace in Fig. \ref{fig2}b) by
$I(V)=I_1+I_2f_E^2+I_3f_E^3$ where we extracted the following parameter: $I_1=$3.21
nA, $I_2=$0.8 nA, $I_3=$1.07 nA, $E_2-E_F^e=$45.86 meV, $E_3-E_F^e=$46.43 meV, $\eta =$0.25, 
$T=$1.6 K. For the fit of the Fano Factor we used Eq. (\ref{eq:Fano}) 
with $\gamma_i=$7.4 ($i=1,2$), $I_1/I_3=$3, and $I_2/I_3=$9.375. The latter current ratio
differs significantly from the values for the current fit. 
In spite of this discrepancy which is not clarified yet
at least the agreement in the peak shape is intriguing and confirms the applicability of
the derived expression (\ref{eq:Fano}) on similar experiments.

\paragraph{Conclusion}

We investigated the bias dependence of shot noise of tunneling through 
self-assembled QDs focusing on the Fano factor modulation observed in our experiment. 
Using a master equation approach we find that the
Fano factor peaks at the current steps are caused by the Pauli exclusion
principle. Neglecting Coulomb interaction we derived an analytical 
expression for the Fano factor which gives good qualitative agreement
with the experiment. The Coulomb interaction $U\approx k_BT$ leads to an additional
dip in the Fano factor.

\paragraph{Acknowledgements}
This work was supported by Deutsche Forschungsgemeinschaft in the framework
of Sfb 296. 
A. N., F. H., and R.J. H. acknowledge financial support from DFG, BMBF, DIP, and TMR.

\end{document}